\documentclass[12pt]{article}
\usepackage{hyperref}
\usepackage[english]{babel}
\usepackage{amsmath}
\usepackage{latexsym}
\usepackage{amssymb}
\usepackage[all]{xy}
\usepackage[dvips]{graphicx}
\usepackage{epsfig}
\usepackage{psfrag}

\numberwithin{equation}{section} \numberwithin{table}{section}
\numberwithin{figure}{section}

\begin{document}


\begin{titlepage}
  \begin{flushright}
  {\small CQUeST-2009-0310}
  \end{flushright}

  \begin{center}

    \vspace{20mm}

    {\LARGE \bf Conductivity and Diffusion Constant in Lifshitz Backgrounds}

    \vspace{10mm}

    Da-Wei Pang$^{\dag}$

    \vspace{5mm}

    {\small \sl $\dag$ Center for Quantum Spacetime, Sogang University}\\
    {\small \sl Seoul 121-742, Korea\\}
    {\small \tt pangdw@sogang.ac.kr}
    \vspace{10mm}

  \end{center}

\begin{abstract}
\baselineskip=18pt We study the DC conductivity and the diffusion
constant for asymptotically Lifshitz black branes in $(d+2)$-
dimensions with arbitrary dynamical exponent $z$. For a solvable
example with $z=2, d=4$, we calculate the real-time correlation
functions, from which we can read off the corresponding conductivity
and diffusion constant. For black branes with arbitrary $z$ and $d$,
we work out the conductivity and obtain the diffusion constant by
making use of the Einstein relation. All the results agree with
those obtained via the membrane paradigm.
\end{abstract}
\setcounter{page}{0}
\end{titlepage}

\pagestyle{plain} \baselineskip=19pt

\tableofcontents

\section{Introduction}
Inspired by condensed matter physics, the investigations on
non-relativistic AdS/CFT correspondence have been accelerated
enormously. Such investigations may open a new window for studying
physical systems in the real world, as many condensed matter systems
are strongly coupled, which can be dealt with in the dual weakly
coupled gravity side via the AdS/CFT correspondence. Several nice
reviews are given in~\cite{Hartnoll:2009sz}.

In many condensed matter systems near a critical point, there exist
field theories with anisotropic scaling symmetry, which means that
the temporal and spatial coordinates scale in different ways,
\begin{equation}
t~\rightarrow~\lambda^{z}t,~~~~~x^{i}~\rightarrow~\lambda x^{i},
\end{equation}
where $z$ is called the `dynamical exponent'. By now, there have
been two main concrete examples of the gravity duals of
non-relativistic field theories. One is the Schr\"{o}dinger case,
which was proposed in~\cite{Son:2008ye, Balasubramanian:2008dm} and
the finite temperature generalizations were investigated
in~\cite{Herzog:2008wg, Maldacena:2008wh, Adams:2008wt}. The other
is the Lifshitz case, which was obtained in~\cite{Kachru:2008yh}.
For general $(d+2)$-dimensional spacetime, the dual geometry of
Lifshitz fixed points is given by
\begin{equation}
ds^{2}=L^{2}(-r^{2z}dt^{2}+\frac{dr^{2}}{r^{2}}+r^{2}d\vec{x}^2),
\end{equation}
where $d\vec{x}^2=dx^{2}_{1}+\cdots+dx^{2}_{d}$. The scale
transformation acts as
\begin{equation}
t~\rightarrow~\lambda^{z}t,~~~x~\rightarrow~\lambda
x,~~~r~\rightarrow~\frac{r}{\lambda}.
\end{equation}
When $z=1$, it reduces to the usual AdS metric.

Unlike the Schr\"{o}dinger case, it is difficult to obtain analytic
black hole solutions in Lifshitz spacetimes. Four dimensional black
hole solutions with $z=2$ were investigated
in~\cite{Danielsson:2009gi} via numerical methods. Lifshitz
topological holes were studied in~\cite{Mann:2009yx} where exact
solutions were found in certain specific examples. Black holes in
asymptotically Lifshitz spacetimes with arbitrary critical exponent
and the corresponding thermodynamic behavior were studied
in~\cite{Bertoldi:2009vn, Bertoldi:2009dt}. Another model of
non-relativistic holography was proposed in~\cite{Taylor:2008tg}
where exact Lifshitz black hole solutions were obtained. Some
properties of such Lifshitz black holes were studied
in~\cite{Pang:2009ad}. Furthermore, certain string theory duals of
Lifshitz-like fixed points were invesitigated
in~\cite{Azeyanagi:2009pr} and some no-go theorems for string duals
of non-relativistic Lifshitz-like theories were proposed
in~\cite{Li:2009pf}. For other recent work on Lifshitz black holes
see~\cite{others}. It should be emphasized that the Lifshitz black
branes arise as the near-horizon geometry of certain holographic
superconductors~\cite{Gubser:2009cg} as well as of charged dilaton
black branes in $AdS_{4}$~\cite{Goldstein:2009cv}. Distinct string
theory realizations of Lifshitz geometries were proposed quite
recently in~\cite{Hartnoll:2009ns}, where a holographic model
building approach to `strange metallic' phenomenology was initiated.

In this paper we focus on the DC conductivity and diffusion constant
in Lifshitz black brane backgrounds, based on the solution obtained
in~\cite{Taylor:2008tg}.\footnote{Such a solution was firstly
obtained in~\cite{Koroteev:2007yp}.} For the relativistic
counterparts, it has been verified in~\cite{Kovtun:2008kx} that the
electrical conductivity and charge susceptibility of a class of
critical models in $d\geq2+1$ dimensions are fixed by the central
charge in a universal manner. The Weyl corrections to the
conductivity and the diffusion constant were studied
in~\cite{Ritz:2008kh}. For the Lifshitz case, a convenient way to
calculate the diffusion constant is to apply the formula derived
from the membrane paradigm~\cite{Kovtun:2003wp, Iqbal:2008by}.
However, when we apply such a formula to calculate the ratio of
shear viscosity over entropy density $\eta/s$, the result has a
non-trivial dependence on the dynamical exponent $z$. When $z=1$, it
reproduces the well-known result $1/4\pi$.

Therefore, one may wonder that if we can still apply the formula
derived from the membrane paradigm to calculate the diffusion
constant in Lifshitz backgrounds. To answer this question, we obtain
the retarded Minkowskian correlation functions for a specific
example $z=2, d=4$, following the standard
approach~\cite{Son:2002sd, Policastro:2002se, Herzog:2002fn}. The
reason why we choose this example is simply that it is tractable.
For arbitrary parameters $z$ and $d$, it is quite difficult to solve
the corresponding differential equations to compute the retarded
Minkowskian correlation functions. Instead we work out the DC
conductivity first and obtain the diffusion constant by making use
of the Einstein relation $D=\sigma/\chi$, where $\chi$ is the charge
susceptibility. Our results exhibit precise agreement with those
obtained via the membrane paradigm approach.

The remaining part of the paper is organized as follows: In Section
2 we review the basic properties of the Lifshitz black branes and
the formulae derived via the membrane paradigm. In Section 3 we
calculate the charge susceptibility $\chi$ for arbitrary $z$ and $d$
and discuss the definition of electric conductivity. In Section 4 we
calculate the Minkowskian correlation functions for a specific
example $z=2, d=4$ and we obtain the diffusion constant and DC
conductivity which can be read off from the corresponding
correlators. In Section 5 we consider the general case with
arbitrary $z$ and $d$ and obtain the DC conductivity first, while
the diffusion constant can be derived by the Einstein relation. All
the results agree with those calculated via the membrane paradigm
approach. Summary and discussion will be given in the final section.

\section{Asymptotically Lifshitz black branes and the membrane paradigm}
In this section we provide some backgrounds which are necessary for
further discussions. It has been observed in~\cite{Taylor:2008tg}
that the effective action in $(d+2)$-dimensional spacetime
\begin{equation}
S=\frac{1}{16\pi G_{d+2}}\int
d^{d+2}x\sqrt{-g}[R-2\Lambda-\frac{1}{2}\partial_{\mu}\phi\partial^{\mu}\phi
-\frac{1}{4}e^{\lambda\phi}\mathcal{F}_{\mu\nu}\mathcal{F}^{\mu\nu}],
\end{equation}
where $\Lambda$ is the cosmological constant and the matter fields
are a massless scalar and an abelian gauge field, admits the
following Lifshitz black branes with arbitrary $z$ as solutions to
the equations of motion
\begin{eqnarray}
\label{2eq2} &
&ds^{2}=L^{2}[-r^{2z}f(r)dt^{2}+\frac{dr^{2}}{r^{2}f(r)}+r^{2}\sum\limits^{d}_{i=1}
dx^{2}_{i}],\nonumber\\
& &f(r)=1-\frac{r_{0}^{z+d}}{r^{z+d}},~~~e^{\lambda\phi}=r^{-2d},~~~\lambda^{2}=\frac{2d}{z-1},\nonumber\\
&
&\mathcal{F}_{rt}=q_{0}r^{z+d-1},~~~q_{0}^{2}=2L^{2}(z-1)(z+d),\nonumber\\
& &\Lambda=-\frac{(z+d-1)(z+d)}{2L^{2}},
\end{eqnarray}
where the horizon locates at $r=r_{0}$. The temperature and entropy
are given by
\begin{equation}
\label{2eq3} T=\frac{z+d}{4\pi}r^{z}_{0},~~~S_{\rm
BH}=\frac{L^{d}V_{d}}{4G_{d+2}}r^{d}_{0},
\end{equation}
where $V_{d}$ denotes the volume of the $d-$dimensional spatial
directions. When $f(r)=1$ with other field configurations remaining
invariant, the zero temperature background is still a solution to
the equations of motion. However, due to the non-trivial profile of
the scalar field, it cannot be seen as the dual gravity description
of the Lifshitz fixed points.

When $z=1$, the solution turns out to be
\begin{eqnarray}
&
&ds^{2}=L^{2}[-r^{2}f(r)dt^{2}+\frac{dr^{2}}{r^{2}f(r)}+r^{2}\sum\limits^{d}_{i=1}
dx^{2}_{i}],\nonumber\\
& &f(r)=1-\frac{r_{0}^{d+1}}{r^{d+1}},~~~\phi=\phi_{0}={\rm
const},\nonumber\\
& &\mathcal{F}_{rt}=0,~~~\Lambda=-\frac{d(d+1)}{2L^{2}},
\end{eqnarray}
which is just the ordinary Schwarzschild-AdS black brane solution.
Therefore the solution cannot be treated as charged under the gauge
field $\mathcal{F}_{rt}$. The scalar field $\phi$ and the gauge
field $\mathcal{F}_{rt}$ simply play the role of modifying the
asymptotic geometry from AdS to Lifshitz.

For general $(p+2)$-dimensional metric
\begin{equation}
ds^{2}=g_{00}(r)dt^{2}+g_{rr}(r)dr^{2}+g_{xx}(r)\sum\limits^{p}_{i=1}dx^{2}_{i},
\end{equation}
the charge diffusion constant is given by the following formula
\begin{equation}
\label{2eq6}
D=\frac{\sqrt{-g(r_{0})}}{g_{xx}(r_{0})\sqrt{-g_{00}(r_{0})g_{rr}(r_{0})}}
\int^{\infty}_{r_{0}}dr\frac{-g_{00}(r)g_{rr}(r)}{\sqrt{-g(r)}},
\end{equation}
which is derived from the membrane paradigm~\cite{Kovtun:2003wp,
Iqbal:2008by}. Substituting the metric in~(\ref{2eq2})
into~(\ref{2eq6}), we obtain
\begin{equation}
\label{2eq7} D=\frac{1}{d-z}r^{z-2}_{0},
\end{equation}
where we have assumed that $z<d$ to ensure the convergence of the
integral. We can rewrite $D$ in terms of the temperature
via~(\ref{2eq3})
\begin{equation}
D=\frac{1}{d-z}(\frac{4\pi}{z+d})^{1-\frac{2}{z}}T^{1-\frac{2}{z}}.
\end{equation}
It turns out that the charge diffusion constant $D$ has a
non-trivial dependence on both the temperature $T$ and the dynamical
exponent $z$. In particular, when $z=1$, the above result reduces to
\begin{equation}
D=\frac{1}{4\pi T}\frac{d+1}{d-1},
\end{equation}
which agrees with the result obtained in~\cite{Kovtun:2008kx}.

However, for gravitational perturbations, the shear mode damping
constant $\mathcal{D}$ is given by~\cite{Kovtun:2003wp,
Iqbal:2008by}
\begin{equation}
\label{2eq10}
\mathcal{D}=\frac{\sqrt{-g(r_{0})}}{\sqrt{-g_{00}(r_{0})g_{rr}(r_{0})}}\int^{\infty}_{r_{0}}
dr\frac{-g_{00}(r)g_{rr}(r)}{g_{xx}(r)\sqrt{-g(r)}}.
\end{equation}
Note that the shear mode damping constant
$\mathcal{D}=\eta/(\epsilon+P)$, where $\eta, \epsilon$ and $P$
denote the shear viscosity, the energy density and the pressure
respectively. We can arrive at the following formula for $\eta/s$ by
using the first law of thermodynamics $\epsilon+P=Ts$
\begin{equation}
\frac{\eta}{s}=T\frac{\sqrt{-g(r_{0})}}{\sqrt{-g_{00}(r_{0})g_{rr}(r_{0})}}\int^{\infty}_{r_{0}}
dr\frac{-g_{00}(r)g_{rr}(r)}{g_{xx}(r)\sqrt{-g(r)}}.
\end{equation}
Substituting the metric in~(\ref{2eq2}) and the temperature
in~(\ref{2eq3}), we obtain
\begin{equation}
\frac{\eta}{s}=\frac{1}{4\pi}\frac{z+d}{d+2-z}r^{2z-2}_{0},
\end{equation}
where the integral is convergent if we still assume $z<d$.

Although the above result can reproduce the well known KSS bound
$1/4\pi$ when $z=1$, the non-trivial dependence on $r_{0}$ for
general $z\neq1$ looks curious. Furthermore, it has been verified
in~\cite{Pang:2009ad} that $\eta/s$ is still $1/4\pi$ for such
Lifshitz black branes. This discrepancy may lead one to suspect the
validity of~(\ref{2eq6}) for Lifshitz black branes. In the
subsequent sections we will see that~(\ref{2eq6}) is still valid for
Lifshitz black branes with general $z$ and $d$.

\section{Charge susceptibility and conductivity}
In the context of relativistic AdS/CFT correspondence, an
equilibrium state at finite temperature and charge density is
described by the Reissner-N\"{o}rdstrom-AdS black holes. For the
Lifshitz case, by now we still do not know the charged solutions
which are dual to equilibrium states with finite temperature and
density. However, to find the charge susceptibility, we need the
relation between the charge density $\rho$ and the chemical
potential $\mu$ to linear order in $\mu$. Then it is sufficient to
choose the uncharged Lifshitz black brane~(\ref{2eq2}) as the
background and to treat the Maxwell action
\begin{equation}
S_{F}=-\frac{1}{4g^{2}_{d+2}}\int
d^{d+2}x\sqrt{-g}F_{\mu\nu}F^{\mu\nu}
\end{equation}
as perturbations. Here $g_{d+2}$ denotes the coupling constant. The
Maxwell equation
$$\frac{1}{\sqrt{-g}}\partial_{\mu}(\sqrt{-g}F^{\mu\nu})=0$$
gives
\begin{equation}
F_{rt}=\frac{1}{r^{d+1-z}}.
\end{equation}
In order to make sure that the gauge potential $A_{t}\rightarrow0$
as $r\rightarrow\infty$, we require
\begin{equation}
z-d\leq-1.
\end{equation}
Notice that this condition also ensures that the integrals
in~(\ref{2eq6}) and~(\ref{2eq10}) are convergent.

The background gauge field is given by
\begin{equation}
A_{t}=\mu-\frac{C}{r^{d-z}},
\end{equation}
where $\mu$ denotes the chemical potential and the constant $C$ is
related to the charge density. The chemical potential $\mu$ is fixed
by the condition that $A_{t}$ vanishes at the horizon $r=r_{0}$,
which gives
$$C=\mu r^{d-z}_{0}.$$
The charge density $\rho$ can be determined by
\begin{equation}
\rho=\frac{\delta\mathcal{L}}{\delta(\partial_{r}A_{t})}
=\frac{L^{d-2}}{g^{2}_{d+2}}(d-z)\mu r^{d-z}_{0}.
\end{equation}
Finally, the charge susceptibility is fixed by the formula
$\rho=\chi\mu$,
\begin{equation}
\label{3eq6} \chi=\frac{L^{d-2}}{g^{2}_{d+2}}(d-z)r^{d-z}_{0}.
\end{equation}

We shall define the electrical conductivity following the approach
proposed in~\cite{Kovtun:2008kx}. Since the dual gravity
descriptions typically do not have dynamical $U(1)$ gauge fields, we
have to clarify the meaning of conductivity first. We imagine
gauging a global $U(1)$ symmetry of the theory with a small coupling
$e$, and work to leading order in $e$. The electrical conductivity
is defined with respect to this $U(1)$ gauge field. At leading
order, the effects of the gauge field can be neglected and the
electromagnetic response can be determined from the original theory.
Then the conserved current $J_{\mu}$ of the $U(1)$ gauge field is
rescaled as $J_{\mu}~\rightarrow~eJ_{\mu}$ and a factor of $e^{2}$
will appear in both the susceptibility and the conductivity.
Finally, the DC conductivity is defined by
\begin{equation}
\sigma=-e^{2}\lim_{\omega\rightarrow0}\frac{1}{\omega}{\rm
Im}~G^{R}_{ii}(\omega,{\bf q}=0),
\end{equation}
where $G^{R}_{ii}(\omega, {\bf q})$ is the retarded real-time
current-current correlation function. For simplicity, we will set
$e^{2}=1$ in the subsequent discussions.
\section{A tractable example: $z=2, d=4$}
In this section we study a concrete example, i.e.$z=2, d=4$ Lifshitz
black brane, following the standard procedures proposed
in~\cite{Policastro:2002se, Herzog:2002fn}. The main reason why we
study this example is simply that it is tractable. We have known in
Section 2 that $z$ and $d$ should satisfy $z-d\leq-1$ such that the
gauge potential has an appropriate asymptotic behavior. Thus in four
dimensions $z>1$ is not allowed while in five dimensions the
corresponding differential equations are difficult to solve. We will
obtain the retarded Minkowskian correlation functions for $z=2,d=4$
case and read off the charge diffusion constant and DC conductivity
from the correlation functions. The DC conductivity can also be
determined from the real-time current-current correlation function
and the charge diffusion constant can be obtained via the Einstein
relation $D=\sigma/\chi$. We will find that the results given by
these two approaches agree.

The main steps for calculating retarded Minkowskian correlation
functions can be summarized as follows~\cite{Son:2002sd,
Policastro:2002se}:
\begin{enumerate}
\item[$(1)$]Extracting the function $B(u)$ staying in front of the
kinetic term $(\partial_{\mu}\phi)^2$ from the action for a certain
field $\phi$,
\begin{equation}
S=\frac{1}{2}\int du d^{d+1}xB(u)(\partial_{u}\phi)^2+\cdots,
\end{equation}
where $u$ denotes the radial coordinate and the boundary locates at
$u=0$.
\item[$(2)$]Solving the linearized equation for $\phi$ and
expressing the bulk field $\phi$ via its boundary value $\phi_{0}$,
\begin{equation}
\phi(u,q)=f_{q}(u)\phi_{0}(q).
\end{equation}
In Minkowski space one has to specify the boundary condition at the
horizon as well as that at the boundary $u=0$. We choose the
incoming-wave boundary condition for all Fourier components
$\phi_{q}$ with timelike $q$ and require regularity for spacelike
$q$'s.
\item[$(3)$]Then the retarded Minkowskian correlation function is
\begin{equation}
\label{4eq3} G^{R}(q)=B(u)f_{-q}(u)\partial_{u}f_{q}(u)|_{u=0}.
\end{equation}
\end{enumerate}

Let us rewrite the six-dimensional $z=2$ Lifshitz black brane as
\begin{equation}
ds^{2}=L^{2}[-r^{4}f(r)dt^{2}+\frac{dr^{2}}{r^{2}f(r)}+r^{2}\sum\limits^{4}_{i=1}dx^{2}_{i}],
~~~f(r)=1-\frac{r^{6}_{0}}{r^{6}},
\end{equation}
where the Hawking temperature is given by
\begin{equation}
T=\frac{6r^{2}_{0}}{4\pi}.
\end{equation}
Introducing the new radial coordinate $u=r^{2}_{0}/r^{2}$, the black
brane metric turns out to be
\begin{equation}
ds^{2}=L^{2}[-\frac{r^{4}_{0}}{u^{2}}f(u)dt^{2}+\frac{du^{2}}{4u^{2}f(u)}
+\frac{r^{2}_{0}}{u}\sum\limits^{4}_{i=1}dx^{2}_{i}],~~~f(u)=1-u^{3},
\end{equation}
where the boundary locates at $u=0$ and the horizon locates at
$u=1$.

The subsequent calculations closely follow~\cite{Policastro:2002se}.
We choose the radial gauge $A_{u}=0$ and Fourier decomposition
\begin{equation}
A_{\mu}=\int\frac{d^{d+1}q}{(2\pi)^{d+1}}e^{-i\omega t+i{\bf
q}\cdot{\bf x}}A_{\mu}(q,u).
\end{equation}
We further fix the spatial momentum by $q_{4}=q$ and $q_{\beta}=0$
for $\beta=1,2,3$. The Maxwell equations
$1/\sqrt{-g}\partial_{\mu}(\sqrt{-g}F^{\mu\nu})=0$ can be rewritten
as
\begin{equation}
\label{4eq8} A^{\prime\prime}_{t}-\frac{1}{4r^{2}_{0}uf(u)}(\omega
qA_{x}+q^{2}A_{t})=0,
\end{equation}
\begin{equation}
\label{4eq9}\omega
A^{\prime}_{t}+qr^{2}_{0}\frac{f(u)}{u}A^{\prime}_{x}=0,
\end{equation}
\begin{equation}
\label{4eq10}\partial_{u}(\frac{f(u)}{u}A^{\prime}_{x})
+\frac{1}{4r^{4}_{0}uf(u)}(\omega^{2}A_{x}+\omega qA_{t})=0,
\end{equation}
\begin{equation}
\label{4eq11}\partial_{u}(\frac{f(u)}{u}A^{\prime}_{\beta})+
\frac{\omega^{2}}{4r^{4}_{0}uf(u)}A_{\beta}-\frac{q^{2}}{4r^{2}_{0}u^{2}}A_{\beta}=0,
\end{equation}
where we have denoted $x_{4}\equiv x$ and the prime stands for
derivative with respect to $u$. Notice that all the equations are
not independent, as we can obtain~(\ref{4eq10}) by
combining~(\ref{4eq8}) and~(\ref{4eq9}).

The following calculations are straightforward. First we can arrive
at the following equation for $A_{t}$ via~(\ref{4eq8})
and~(\ref{4eq9})
\begin{equation}
\label{4eq12}
A^{\prime\prime\prime}_{t}+\frac{f^{\prime}(u)}{f(u)}A^{\prime\prime}_{t}
+\frac{1}{u}A^{\prime\prime}_{t}+\frac{\omega^{2}}{4r^{4}_{0}f^{2}(u)}A^{\prime}_{t}
-\frac{q^{2}}{4r^{2}_{0}uf(u)}A^{\prime}_{t}=0.
\end{equation}
To determine the behavior of $A^{\prime}_{t}$ near the singular
point $u=1$, we set $A^{\prime}_{t}=(1-u)^{\alpha}F(u)$, where
$F(u)$ is a regular function and substitute it into~(\ref{4eq12}).
One can find that $\alpha^{2}=-\omega^{2}/36r^{4}_{0}$ and the
incoming-wave boundary condition determines
\begin{equation}
\alpha=-\frac{i\omega}{6r^{2}_{0}}=-\frac{i\omega}{4\pi T}.
\end{equation}

Next we will solve the differential equation perturbatively by
setting
\begin{equation}
F(u)=F_{0}(u)+\omega F_{1}(u)+q^{2}G_{1}(u)+\mathcal{O}(\omega^{2},
\omega q, q^{2}).
\end{equation}
The equation for $F_{0}(u)$ is given by
\begin{equation}
F^{\prime\prime}_{0}(u)+\frac{f^{\prime}(u)}{f(u)}F^{\prime}_{0}(u)
+\frac{1}{u}F^{\prime}_{0}(u)=0,
\end{equation}
whose solution is simply $F_{0}(u)=C$, where $C$ is a constant. Then
the equations for $F_{1}(u)$ and $G_{1}(u)$ can be simplified
considerably,
\begin{equation}
F^{\prime\prime}_{1}(u)+[\frac{f^{\prime}(u)}{f(u)}+\frac{1}{u}]F^{\prime}_{1}(u)
+\frac{iC}{4\pi
T}[\frac{f^{\prime}(u)}{(1-u)f(u)}+\frac{1}{u(1-u)^{2}}]=0,
\end{equation}
\begin{equation}
G^{\prime\prime}_{1}(u)+[\frac{f^{\prime}(u)}{f(u)}+\frac{1}{u}]G^{\prime}_{1}(u)
-\frac{C}{4r^{2}_{0}uf(u)}=0.
\end{equation}
The solutions are given by
\begin{equation}
F_{1}(u)=\frac{iC}{4\pi T}[3\ln u-\ln\frac{1}{3}(1+u+u^{2})],
\end{equation}
\begin{equation}
G_{1}(u)=\frac{C}{4r^{2}_{0}}[-\ln
u+\frac{1}{2}\ln\frac{1}{3}(1+u+u^{2})+\frac{\sqrt{3}}{3}
(\tan^{-1}\frac{2u+1}{\sqrt{3}}-\frac{\pi}{3})],
\end{equation}
where we have fixed the integration constants by requiring that
$F_{1}(u), G_{1}(u)$ are regular at the horizon $u=1$ and
$F_{1}(1)=G_{1}(1)=0$.

We can also obtain the following relation between $A_{x}$ and
$A_{t}$ from~(\ref{4eq8}),
\begin{equation}
A_{x}=\frac{4r^{2}_{0}uf(u)}{\omega
q}A^{\prime\prime}_{t}-\frac{q}{\omega}A_{t}.
\end{equation}
Then the integration constant $C$ can be fixed in terms of the
boundary value of the fields $A^{0}_{t}, A^{0}_{x}$,
\begin{equation}
C=\frac{q\omega A^{0}_{x}+q^{2}A^{0}_{t}}{2i\omega-q^{2}}.
\end{equation}
We will see that the pole in $C$ is the same pole that appears in
the retarded Green's functions.

The Maxwell action can be recast as
\begin{eqnarray}
S_{F}&=&-\frac{1}{2g^{2}_{6}}\int
dud^{5}x\sqrt{-g}g^{uu}g^{ij}\partial_{u}A_{i}\partial_{u}A_{j}\nonumber\\
&=&\frac{L^{2}r^{2}_{0}}{g^{2}_{6}}\int
dud^{5}x[A^{\prime2}_{t}-r^{2}_{0}\frac{f(u)}{u}A^{\prime2}_{i}].
\end{eqnarray}
Then we can extract the function $B(u)$,
\begin{equation}
B_{t}(u)=\frac{2L^{2}r^{2}_{0}}{g^{2}_{6}},~~~B_{i}(u)=-\frac{2L^{2}r^{4}_{0}}
{g^{2}_{6}}\frac{f(u)}{u}.
\end{equation}
We can easily obtain the real-time correlation functions following
the standard prescriptions
\begin{equation}
G^{R}_{tt}=\frac{L^{2}r^{2}_{0}}{g^{2}_{6}}\frac{q^{2}}{i\omega-\frac{1}{2}q^{2}},
\end{equation}
\begin{equation}
G^{R}_{xx}=\frac{L^{2}r^{2}_{0}}{g^{2}_{6}}\frac{\omega^{2}}{i\omega-\frac{1}{2}q^{2}},
\end{equation}
\begin{equation}
G^{R}_{tx}=G^{R}_{xt}=-\frac{L^{2}r^{2}_{0}}{g^{2}_{6}}\frac{\omega
q}{i\omega-\frac{1}{2}q^{2}}.
\end{equation}
By making use of the general form of the retarded correlation
function
$$G^{R}_{tt}=\frac{\chi Dq^{2}}{i\omega-Dq^{2}}$$
and the Einstein relation $\sigma=\chi D$, we can read off the
diffusion constant and the conductivity
\begin{equation}
D=\frac{1}{2},~~~\sigma=\frac{L^{2}r^{2}_{0}}{g^{2}_{6}}.
\end{equation}
Comparing these results with~(\ref{2eq7}), we can find precise
agreement when $z=2, d=4$.

Unfortunately, the differential equation for $A_{\beta}$ is
difficult to solve, so we cannot obtain the corresponding retarded
correlation functions. However, to obtain the conductivity we just
need to calculate $G^{R}_{\beta\beta}(\omega, q=0)$. Then the
equation for $A_{\beta}$ is given by
\begin{equation}
\partial_{u}(\frac{f(u)}{u}A^{\prime}_{\beta})+
\frac{\omega^{2}}{4r^{4}_{0}uf(u)}A_{\beta}=0.
\end{equation}
Similarly, we assume $A_{\beta}=(1-u)^{\gamma}H(u)$, where $H(u)$ is
a regular function. The incoming-wave boundary condition still
forces us to choose $\gamma=-i\omega/4\pi T$. Next we solve for
$H(u)$ perturbatively,
\begin{equation}
H(u)=H_{0}(u)+\omega H_{1}(u)+\mathcal{O}(\omega^{2}).
\end{equation}
The equation for $H_{0}$ is given by
\begin{equation}
H^{\prime\prime}_{0}(u)+\frac{f^{\prime}(u)}{f(u)}H^{\prime}_{0}(u)-
\frac{1}{u}H^{\prime}_{0}(u)=0,
\end{equation}
and the solution is simply
\begin{equation}
H_{0}(u)=H_{0}={\rm const}.
\end{equation}
The equation for $H_{1}(u)$ takes the following simple form after
fixing $H_{0}$,
\begin{equation}
H^{\prime\prime}_{1}(u)+[\frac{f^{\prime}(u)}{f(u)}-\frac{1}{u}]H^{\prime}_{1}(u)
+\frac{iH_{0}}{4\pi
T}[\frac{f^{\prime}(u)}{(1-u)f(u)}+\frac{2u-1}{u(1-u)^{2}}]=0,
\end{equation}
whose solution is
\begin{equation}
H_{1}(u)=\frac{iH_{0}}{4\pi T}[\frac{1}{2}\ln
\frac{1}{3}(1+u+u^{2})-\sqrt{3}(\tan^{-1}\frac{2u+1}{\sqrt{3}}-\frac{\pi}{3})].
\end{equation}
Here the integration constants have been fixed such that $H_{1}(u)$
is regular at the horizon $u=1$ and $H_{1}(1)=0$.

The constant $H_{0}$ is fixed by
\begin{equation}
H_{0}=\frac{A^{0}_{\beta}}{1+\frac{i\omega}{4\pi
T}(\frac{\pi}{2\sqrt{3}}-\ln3)},
\end{equation}
where $A^{0}_{\beta}$ stands for the boundary value of $A_{\beta}$.
Finally, from~(\ref{4eq3}) we can obtain
\begin{equation}
G^{R}_{\beta\beta}(\omega,
q=0)=-\frac{L^{2}r^{2}_{0}}{g^{2}_{6}}i\omega.
\end{equation}
Then the DC conductivity is given by
\begin{equation}
\sigma=-\lim_{\omega\rightarrow0}\frac{1}{\omega}{\rm Im}
G^{R}_{\beta\beta}(\omega, q=0)=\frac{L^{2}r^{2}_{0}}{g^{2}_{6}}.
\end{equation}
We can determine the diffusion constant by using~(\ref{3eq6}) and
the Einstein relation,
\begin{equation}
D=\frac{\sigma}{\chi}=\frac{1}{2}.
\end{equation}
Therefore we find precise agreement with~(\ref{2eq7}) for $z=2, d=4$
once again.
\section{Conductivity and diffusion constant for arbitrary $z$ and $d$}
In this section we will calculate the DC conductivity and diffusion
constant for arbitrary $z$ and $d$. Generally speaking, the
differential equations for arbitrary $z$ and $d$ are difficult to
solve thus we can hardly read off the conductivity and diffusion
constant from the retarded correlation functions straightforwardly.
However, as stated in the previous section. We can obtain the
conductivity first by evaluating the retarded correlation function
$G^{R}_{ii}(\omega, q=0)$, which is always tractable. The diffusion
constant can be obtained by the Einstein relation $D=\sigma/\chi$.
We will see that the results for general $z$ and $d$ agree with
those obtained via the membrane paradigm.

Recall the Lifshitz black brane solution for arbitrary $z$ and $d$
$$ds^{2}=L^{2}[-r^{2z}f(r)dt^{2}+\frac{dr^{2}}{r^{2}f(r)}+r^{2}
\sum\limits^{d}_{i=1}dx^{2}_{i}],~~~f(r)=1-\frac{r^{z+d}_{0}}{r^{z+d}}.$$
Consider the coordinate transformation $u=r_{0}/r$, the metric can
be rewritten as
\begin{equation}
ds^{2}=L^{2}[-\frac{r^{2z}_{0}}{u^{2z}}f(u)dt^{2}+\frac{du^{2}}{u^{2}f(u)}
+\frac{r^{2}_{0}}{u^{2}}\sum\limits^{d}_{i=1}dx^{2}_{i}],~~~
f(u)=1-u^{z+d}.
\end{equation}
The horizon still locates at $u=1$ and the boundary locates at
$u=0$.

Since we only need to work out $G^{R}_{ii}(\omega, q=0)$, we can
assume the gauge field to be $A_{i}=A_{i}(u)e^{-i\omega t}$. The
corresponding Maxwell equations can be written as
\begin{equation}
\partial_{u}(\frac{f(u)}{u^{z+d-3}}A^{\prime}_{i}(u))
+\omega^{2}\frac{u^{z-d+1}}{r^{2z}_{0}f(u)}A_{i}(u)=0,
\end{equation}
where the prime denotes derivative with respect to the radial
coordinate $u$. Let $A_{i}(u)=(1-u)^{\delta}A(u)$, where $A(u)$ is a
regular function. Similarly, the regularity at $u=1$ and the
incoming-wave boundary condition still fix $\delta=-i\omega/4\pi T$.

Next we solve for $A(u)$ perturbatively,
\begin{equation}
A(u)=A_{0}(u)+i\omega A_{1}(u)+\mathcal{O}(\omega^{2}).
\end{equation}
The equation for $A_{0}(u)$ is given by
\begin{equation}
A^{\prime\prime}_{0}(u)+\frac{f^{\prime}(u)}{f(u)}A^{\prime}_{0}(u)
-\frac{z+d-3}{u}A^{\prime}_{0}(u)=0,
\end{equation}
whose solution is still simply
\begin{equation}
A_{0}(u)=A_{0}={\rm const}.
\end{equation}
The equation for $A_{1}(u)$ can be written as
\begin{equation}
A^{\prime\prime}_{1}(u)+[\frac{f^{\prime}(u)}{f(u)}-\frac{z+d-3}{u}]
A^{\prime}_{1}(u)+\frac{A_{0}}{4\pi
T}[\frac{1}{(1-u)^{2}}+\frac{f^{\prime}(u)}{f(u)(1-u)}-\frac{z+d-3}{u(1-u)}]=0.
\end{equation}
The solution for $A_{1}(u)$ can be expressed in terms of
hypergeometric functions,
\begin{equation}
A_{1}(u)=\frac{A_{0}}{r^{z}_{0}}(\frac{1}{2u^{2}}-\frac{1}{2u^{2}}
{}_{2}F_{1}[-\frac{2}{z+d},1,1-\frac{2}{z+d},u^{z+d}])+C_{1}-\frac{A_{0}}{4\pi
T}\ln(1-u),
\end{equation}
where we have fixed one integration constant by requiring that
$A_{1}(u)$ is regular at $u=1$ and the other integration constant
$C_{1}$ ensures that $A_{1}(u)$ vanishes at $u=1$. Moreover, $A_{0}$
can be determined in terms of the boundary value $A^{0}_{i}$,
\begin{equation}
A^{0}_{i}=A_{0}+\mathcal{O}(\omega).
\end{equation}

Now let us rewrite the Maxwell action as
\begin{eqnarray}
S_{F}&=&-\frac{1}{2g^{2}_{d+2}}\int
dud^{d+1}x\sqrt{-g}g^{uu}g^{ij}\partial_{u}A_{i}\partial_{u}A_{j}\nonumber\\
&=&-\frac{1}{2g^{2}_{d+2}}\int
dud^{d+1}x\frac{L^{d-2}r^{z+d-2}_{0}}{u^{z+d-3}}f(u)(\partial_{u}A_{i})^{2}.
\end{eqnarray}
Then we can extract the function $B(u)$
\begin{equation}
B(u)=-\frac{L^{d-2}r^{z+d-2}}{g^{2}_{d+2}u^{z+d-3}}f(u).
\end{equation}
Finally, the retarded correlation function is given by
\begin{equation}
G^{R}_{ii}(\omega,
q=0)=-\frac{L^{d-2}r^{d-2}_{0}}{g^{2}_{d+2}}i\omega.
\end{equation}
Then the DC conductivity is determined as
\begin{equation}
\sigma=-\lim_{\omega\rightarrow0}\frac{1}{\omega}{\rm Im}
G^{R}_{ii}(\omega, q=0)=\frac{L^{d-2}r^{d-2}_{0}}{g^{2}_{d+2}}.
\end{equation}
By combining the charge susceptibility~(\ref{3eq6}) and the Einstein
relation, we can arrive at
\begin{equation}
D=\frac{1}{d-z}r^{z-2}_{0},
\end{equation}
which agrees with the result given by the membrane
paradigm~(\ref{2eq7}) for arbitrary $z$ and $d$.
\section{Summary and discussion}
The membrane paradigm provides us an efficient way of investigating
the hydrodynamics of black holes. In particular, the charge
diffusion constant and the ratio of shear viscosity over entropy
density $\eta/s$ can be evaluated in terms of the black hole metric
and temperature. However, when we apply the formula to the Lifshitz
black branes straightforwardly, we find that the well-known result
$\eta/s=1/4\pi$ cannot be reproduced. This leads one to suspect the
validity of the formula for the charge diffusion constant. In this
paper we study the conductivity and the charge diffusion constant
for Lifshitz black branes. We calculate the retarded correlation
functions for a specific, tractable example $z=2, d=4$, following
the standard approach proposed in~\cite{Policastro:2002se}. Then we
read off the charge diffusion constant and the conductivity from the
$tt$-component of the correlators. We also obtain the conductivity
for general $z$ and $d$ by evaluating the corresponding correlation
functions with $q=0$ and the charge diffusion constant via the
Einstein relation. All these results agree with those obtained via
the membrane paradigm. This verifies the validity of the formula for
the charge diffusion constant derived from the membrane paradigm.
However, one may need to find the appropriate formula for $\eta/s$
in the Lifshitz black brane backgrounds.

For the relativistic counterparts, it has been shown
in~\cite{Kovtun:2008kx} that for a large class of critical models in
$d\geq2+1$ dimensions, the electrical conductivity and charge
susceptibility are fixed the central charge in a universal manner. A
universal relation for $\sigma/\chi$ is proposed, which looks
similar to the famous relation for $\eta/s$. It was further shown
in~\cite{Ritz:2008kh} that such a lower bound still held if the Weyl
corrections were considered. However, as pointed out
in~\cite{Kovtun:2008kx}, if $\hbar$ and the speed of light $v$ were
restored, the ratio turned out to be
\begin{equation}
\frac{\sigma}{\chi}\geq\frac{\hbar v^{2}}{4\pi T}{\frac{d}{d-2}}.
\end{equation}
 Such a bound cannot hold in non-relativistic systems as the speed
of light $v\rightarrow\infty$. Therefore we cannot expect a similar
bound for the Lifshitz black branes.

It should be pointed out that comparing the relativistic
counterparts, i.e. RN-AdS black holes, we are still lack of charged
black brane solutions in asymptotically Lifshitz spacetime, which
can be seen as the dual gravity description of field theories at
Lifshitz fixed points at finite temperature and charge density. Once
we obtain such charged black brane solutions, we may investigate
their thermodynamic and hydrodynamic properties, following
e.g.~\cite{Herzog:2007ij, Hartnoll:2007ai, Hartnoll:2007ih,
Hartnoll:2007ip}. Moreover, three distinct string theory
realizations of Lifshitz geometries were outlined
in~\cite{Hartnoll:2009ns}. It would be interesting to study the
corresponding properties of such Lifshitz geometries in the context
of string theory via a top-down approach.

\bigskip \goodbreak \centerline{\bf Acknowledgements}
\noindent DWP would like to thank Rong-Gen Cai, Bom Soo Kim and
Jian-Huang She for helpful discussions. This work was supported by
the National Research Foundation of Korea(NRF) grant funded by the
Korea government(MEST) through the Center for Quantum
Spacetime(CQUeST) of Sogang University with grant number
2005-0049409.



\end{document}